# On-Demand Growth of Semiconductor Heterostructures Guided by Physics-Informed Machine Learning

Chao Shen[1,2], Yuan Li[1,2], Wenkang Zhan[1,2], Shujie Pan[1,3], Fuxin Lin[1], Kaiyao Xin[2,4], Hui Cong[2,5], Chi Xu[2,5], Xiaotian Cheng[6], Ruixiang Liu[6], Zhibo Ni[6], Chaoyuan Jin[6], Bo Xu[1,2], Siming Chen[1,2], Zhongming Wei[2,4], Chunlai Xue[2,5], Zhanguo Wang[1,2], and Chao Zhao[1,2,*]

[1] Laboratory of Solid State Optoelectronics Information Technology, Institute of Semiconductors, Chinese Academy of Sciences, Beijing 100083, China

[2] Center of Materials Science and Optoelectronics Engineering, University of Chinese Academy of Sciences, Beijing 100049, China

[3] HS Photonics Co., Ltd., Xiangjiang Science & Technology Innovation Base, Changsha, Hunan 413000, China

[4] State Key Laboratory of Semiconductor Physics and Chip Technologies, Institute of Semiconductors, Chinese Academy of Sciences, Beijing 100083, China

[5] State Key Laboratory of Optoelectronic Materials and Devices, Institute of Semiconductors, Chinese Academy of Sciences, Beijing 100083, China

[6] College of Information Science and Electronic Engineering, State Key Laboratory of Silicon and Advanced Semiconductor Materials, Zhejiang University, Hangzhou 310027, China

*Email: zhaochao@semi.ac.cn

**Abstract**

Developing tailored semiconductor heterostructures on demand represents a critical capability for addressing the escalating performance demands in electronic and optoelectronic devices. However, traditional fabrication methods remain constrained by simulation-based design and iterative trial-and-error optimization. Here, we introduce SemiEpi, a self-driving platform designed for molecular beam epitaxy (MBE) to perform multi-step semiconductor heterostructure growth through in-situ monitoring and on-the-fly feedback control. By integrating standard MBE reactors, physics-informed machine learning (ML) models, and parameter initialization, SemiEpi identifies optimal initial conditions and proposes experiments for heterostructure growth, eliminating the need for extensive expertise in MBE processes. As a proof of concept, we demonstrate the optimization of high-density InAs quantum dot (QD) growth with a target emission wavelength of 1240 nm, showcasing the power of SemiEpi. We achieve a QD density of $5 \times 10^{10}$ $cm^{-2}$, a 1.6-fold increase in photoluminescence (PL) intensity, and a reduced full width at half maximum (FWHM) of 29.13 meV, leveraging in-situ reflective high-energy electron diffraction monitoring with feedback control for adjusting growth temperatures. Taken together, our results highlight the potential of ML-guided systems to address challenges in multi-step heterostructure growth, facilitate the development of a hardware-independent framework, and enhance process repeatability and stability, even without exhaustive knowledge of growth parameters.

## Introduction

Semiconductor heterostructures are fundamental building blocks in modern electronic and optoelectronic devices. Through precise epitaxial growth, these heterostructures can be customized to achieve desired characteristics such as emission wavelength, carrier mobility, and reflectivity to meet diverse application requirements. For instance, laser diodes require meticulously engineered heterostructures that effectively confine both carriers and photons, ensuring stable emission at target wavelengths.[1] In contrast, high-electron-mobility transistors incorporate heterojunctions as the channel for electrons to enhance device performance.[2] However, the development of such heterostructures involves a complex multi-step growth process requiring careful optimization of numerous parameters, with each epitaxial layer demanding specific customization. Traditionally, this process has heavily relied on trial and error, making it both time-consuming and resource-intensive.[3] Transitioning from material discovery to practical implementation can span decades. As functional demands increase, there is an urgent need for fast, on-demand development of customized semiconductor heterostructures. Achieving such precision has proven to be a significant challenge for many years due to its complex nature.

In recent years, self-driving experiments guided by machine learning (ML) have shown potential to transform research by streamlining traditionally repetitive tasks with minimal human involvement.[4] This approach has been successfully employed in small-molecule synthesis, photocatalytic reactions, and the discovery of organic lasers.[5] Moreover, ML has proven particularly effective in uncovering relationships between dynamic growth processes and final material properties, facilitating the optimization of growth parameters.[6] However, it faces two key limitations. First, it relies heavily on historical growth datasets, which limits its adaptability to unexplored growth scenarios. Second, the inherent "blacktep-box" and data-hungry nature of ML models obscure interpretability. Integrating physical principles into ML frameworks, named as physics-informed ML (PIML), could enhance both generalization and model transparency, yet this remains a persistent challenge.[7]

Furthermore, semiconductor heterostructures present unique challenges due to their inherent complexity, which involves designing the structure through simulation and optimizing multiple interdependent variables, such as growth temperature, flux, and growth rate. Among heterostructures, quantum dot (QD)-based heterostructures are at the forefront of semiconductor research, driven by the growing demand for high-performance and miniaturized devices.[8] However, developing QD heterostructures introduces additional variables, including precise control over QD size, shape, and density to meet device requirements.[9] Moreover, dynamic fluctuations in reactor conditions and batch-to-batch variations in sample holders and substrates can render predetermined growth parameters and programs ineffective.[10] To address these issues, it is essential to establish a self-consistent parameter initialization protocol grounded in the material's intrinsic properties, rather than relying solely on historical growth data.

Reconstruction transitions, typically monitored in situ via reflective high-energy electron diffraction (RHEED), can vary based on the material's inherent physical properties. Among growth parameters, temperature stands out as both easily measurable and critically important for QD growth.[11] The characteristic temperatures associated with RHEED transitions can help identify optimal growth conditions.[12] In this context, we reported the growth of QDs and lasers using ML and in-situ feedback control assisted by RHEED.[13,14,15] However, initial conditions still require human intervention, and methods developed for one molecular beam epitaxy (MBE) reactor may not apply to another, which limits the accessibility of customized heterostructures. A robust platform is essential to address these challenges by integrating PIML models,[16] multi-step processes and enabling real-time optimization. It should prioritize performance-driven heterostructure design, ensure proper parameter initialization, develop predictive models of growth outcomes, and dynamically optimize growth parameters throughout the growth process.

In our work, we introduce SemiEpi, a self-driving platform designed to perform multi-step MBE growth of heterostructures under optimal conditions. It integrates three key innovations: (1) PIML models for QD design, (2) a self-driving method for multi-step heterostructure growth with minimal human involvement, and (3) in-situ monitoring with feedback control for adjusting growth temperatures. Using InAs QDs as a demonstrative heterostructure, SemiEpi employs PIML models to determine the QD size and density based on the desired emission wavelength. The core functionality of SemiEpi involves in situ RHEED videos analysis to monitor surface reconstruction and QD growth. It generates a parameter initialization curve, selects a proper initial growth temperature to ease the burden on the ML models, and fine-tunes the substrate temperature during growth using ML without human intervention. This capability enables the customization of growth conditions for individual heterostructure, marking a significant milestone in establishing a precise growth control scheme and closed-loop experimentation strategies. This way, we can intelligently optimize multi-step growth within a complex parameter space, achieving desired material characteristics previously only attainable through time-consuming, labor-intensive experimentation and human intervention. Our platform represents a significant advancement toward self-driving semiconductor epitaxy, combining physical insights with adaptive ML for unprecedented control over nanoscale material synthesis.

## Results

### Design of SemiEpi

Recent advances in self-driving material synthesis systems have demonstrated the potential to achieve on-demand material properties by leveraging historical experimental data to iteratively optimize growth parameters.[17,18,19,20] A critical yet often overlooked aspect of these systems is proper parameter initialization to minimize deviations during growth and reduce the need for adjustments, improving universality across different systems to achieve optimal outcomes. Although recent studies have proposed theoretical frameworks for parameter initialization and growth outcome prediction using in situ data, these methodologies have yet to be implemented or integrated into dynamic control frameworks.[21,22]

SemiEpi was designed to address the challenge of optimizing complex, multi-step growth processes for demanded material properties. It integrates PIML models that combine physical constraints to ensure robust interpretability while allowing exploration of parameter spaces associated with QD sizes corresponding to specific emission wavelengths (see Fig. 1a). PIML further leverages ML to uncover complex, nonlinear relationships between growth outcomes, such as QD density and size, which can be inferred from RHEED pattern analysis.[23] This approach opens up opportunities to control over materials properties during growth. Once the target QD size linked to the desired emission wavelength is identified, PIML models can predict the corresponding QD density. This methodology transforms complex growth characteristics, often challenging to observe directly, into accessible and measurable parameters.

SemiEpi's heterostructure growth process is governed by three sequential operational modules: (1) substrate deoxidation, (2) parameter initialization, and (3) material growth based on in-situ characterization results and ML models called "Initialization Model" and "Growth Models". It facilitates continuous in-situ monitoring, self-optimization, and on-the-fly feedback control to grow specific structures. The process initiates with the substrate deoxidation module, which performs critical surface preparation by thermally removing native oxides to ensure a fresh growth front. During this phase, the substrate temperature gradually increased until the deoxidation feature is observed at ~580 °C through in-situ RHEED, and the deoxidation temperature is subsequently recorded (see Fig. 1b).

The parameter initialization module employs RHEED to monitor real-time surface reconstruction states using the Initialization Model while simultaneously recording the corresponding thermocouple temperatures. During substrate temperature modulation, surface energy and stress variations drive atomic rearrangement, enabling observation of distinct reconstruction states.[24] Each state transition occurs at a characteristic temperature: for example, the As cap desorbs from the GaAs surface at a theoretical temperature of 350 °C as surface energy decreases with increasing temperature.[25] Further heating induces sequential reconstruction transitions from c(4×4) to (2×4) at ~510 °C, followed by (2×4) to (n×6) at ~620 °C (see Fig. 1c).[26] By correlating measured thermocouple readings with these theoretically established transition temperatures, SemiEpi constructs a robust parameter initialization curve (see Fig. 1d). This enables precise alignment of actual growth temperatures with theoretical optima. In this way, SemiEpi provides inherent adaptability across different MBE systems, minimizing the need for reactor-specific parameter adjustments in subsequent processes.

Following parameter initialization, the 100-nm GaAs is re-grown at a theoretical temperature of 600 °C based on the parameter initialization results. The substrate is subsequently cooled to a theoretical temperature of 490 °C for sequential growth of buried InAs QDs, followed by a 10-nm GaAs layer. The temperature is then raised to a theoretical temperature of 600 °C to grow an additional 100-nm GaAs. Finally, the substrate is cooled to the temperature used for buried InAs QDs growth, and the surface InAs QDs are grown starting from this temperature. The material growth module performs heterostructure growth; it records real-time data, including temperature and shutter status, which is pre-processed by the software and then fed into the Growth Models selected depending on the input data type for further processing. Hardware acceleration, controlled by the software, enhances data processing efficiency (see Fig. 1e). The module also monitors the material's status during growth, employs Growth Model to identify conditions, predicts growth outcomes, and maps them to a parameter optimization table for further optimization. This process enables the software to send instructions, thereby ensuring dynamic feedback control of the growth parameters. The Growth Models efficiently analyze a large amount of in-situ monitoring data. Operating in a closed-loop manner, the module feeds real-time growth information back into the model to generate insights that iteratively guide parameter optimization until the desired conditions are achieved.

Combining PIML with on-the-fly feedback control, SemiEpi automates execution and analysis. This approach allows for dynamic optimization of these parameters within a single experiment, minimizing the need for multiple experimental runs and resulting in more efficient and effective material growth. SemiEpi is a self-driving setup that offers significant advantages over conventional methods by customizing growth conditions for each substrate, ensuring compatibility across different reactors without requiring code modifications. It eliminates the need for extensive semiconductor or MBE process expertise to achieve the best results, making it a valuable platform for use in any epitaxy laboratory regardless of the grower's level of knowledge and experience.

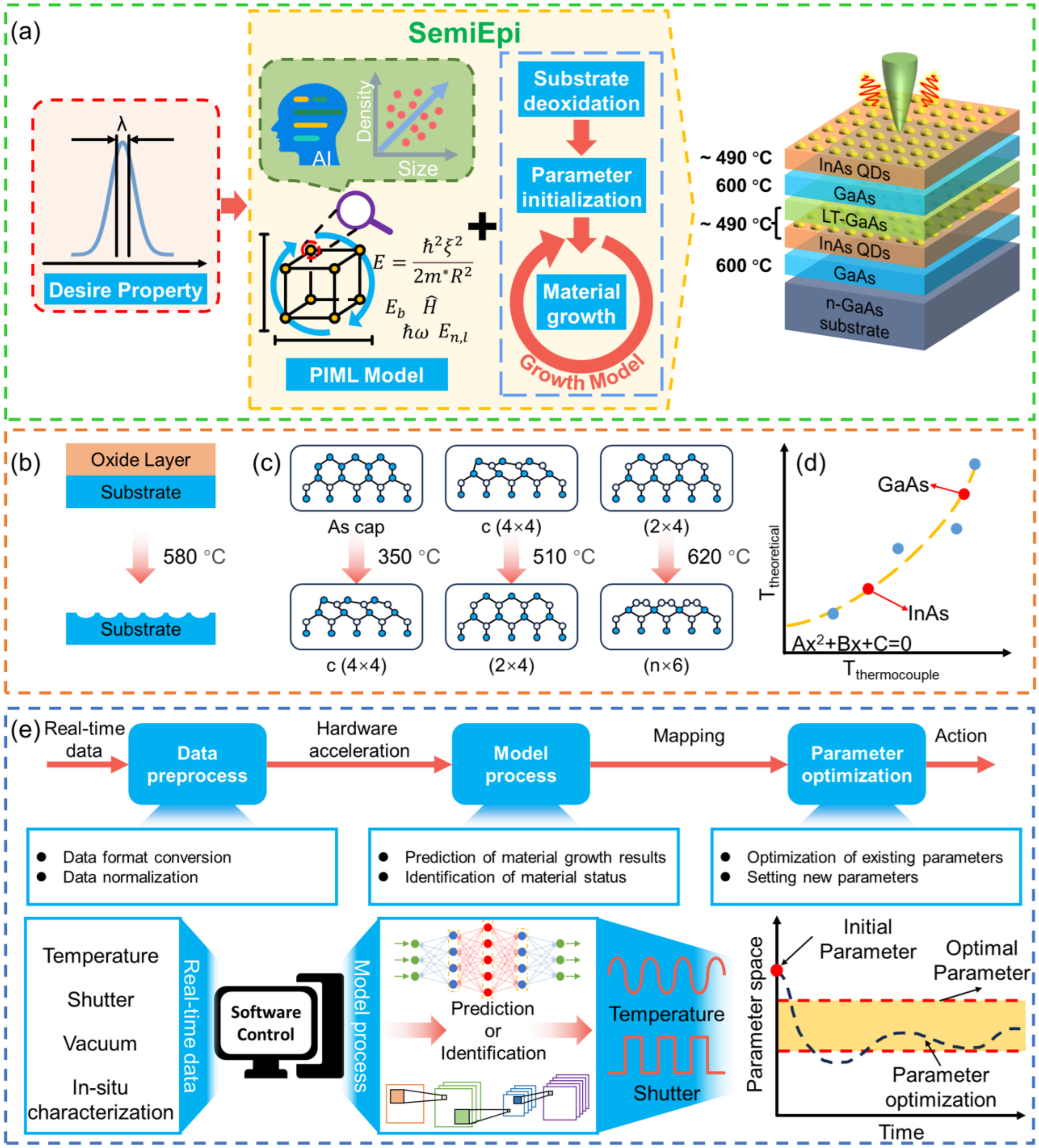

**Fig. 1: Overview of SemiEpi.** (a) SemiEpi's framework. Before SemiEpi, growers defined target properties. SemiEpi then employs a multi-module, sequential workflow to design and grow samples with specified structures. Schematic of (b) substrate deoxidation, (c) surface reconstruction, (d) parameter initialization, and (e) material growth.

## SemiEpi configuration

### Software and Hardware

Existing industrial software functions as isolated, sequential processing systems that maintain a fundamental disconnect between material growth and characterization data. This segregation prevents hinders effective data fusion between material growth and characterization systems, making it difficult to adjust growth parameters in real-time. Additionally, most industrial software prioritizes automating predefined steps rather than intelligently evaluating or adjusting the growth condition based on real-time data.

In our work, SemiEpi was developed using LabVIEW. It utilizes NI-VISA to control parameters for MBE. Specifically, it communicates with address-specific controllers through the Modbus protocol, allowing for the reading and writing of commands to control temperatures. Additionally, it employs direct binary serial commands, where specific bit positions correspond to the states of individual shutters, enabling precise control over different shutters. It also utilizes NI VISION to acquire RHEED video from the fluorescent screen. The acquired data is processed using Python libraries and transferred to the ML model, formatted in ONNX and optimized with TensorRT for faster inference (see Supplementary Information for the platform environment and program interface and the functional interface switch in the program, S1 and S2).

SemiEpi combines a standard MBE reactor, a camera, temperature controllers, shutter controllers, and an in-situ RHEED system to enable monitoring and optimization. It was designed and deployed on a Windows 10 system with an AMD R9 7950X CPU, 64GB of RAM, an NVIDIA 3090 graphics card, and a 2TB solid-state drive. The system is linked to a temperature controller and a shutter controller via USB 2.0 for data exchange. The Modbus protocol enables the connection of multiple temperature controllers in series and precise control of the In and Ga cells using addresses. Additionally, USB 3.0 allows the connection of a camera in a dark room outside the fluorescent screen. Furthermore, the model training and data preprocessing processes used by SemiEpi are also conducted on the system (see Supplementary Information for the deployment environment and hardware wiring scheme, S3).

**PIML model construction and utilization**

SemiEpi begins with PIML models that design a QD heterostructure for an emission wavelength of 1240 nm, suitable for gas sensing applications (see Fig. 2a).[27] The PIML model is implemented in two stages to achieve this design. First, the target wavelength is input into a physics model that simulates QDs with finite potential barriers (see Supplementary Information for the construction of the physics model, S4).[28] Due to fluctuations during growth and material inhomogeneity, QDs emitting the same target wavelength can vary in size. As a result, the model is executed multiple times, and the mean and standard deviation of the results are statistically analyzed. Ultimately, the model predicts that the QDs have a mean diameter of 20.90 nm with a standard deviation (std) of 2.74 nm and a mean height of approximately 5.56 nm with a standard deviation of 0.98 nm.

Second, we estimate QD density ranges from predicted sizes to enable indirect monitoring. Determining the average QD size directly is only feasible after the growth transition from a 2D wetting layer to a 3D island-like structure, and the size evolution typically happens within just a few seconds. This narrow time window limits the opportunity to optimize parameters to achieve the desired QD properties.[29] However, previous studies have shown that RHEED can predict changes in QD density before QD formation.[14] Since QD size is correlated with density, this presents an opportunity to indirectly infer QD size by observing QD density.[14] To this end, the PIML model estimates QD density ranges and values based on QD size. We collected approximately 380 data points from atomic force microscopy (AFM) to create a robust dataset, which facilitates modeling the relationship between QD density and size.

We conducted a comprehensive analysis of the dataset to model the relationship between QD density and size. The feature correlation heatmap revealed a strong positive correlation between "Diameter std" and "Height std", with a coefficient of 0.75 (see Fig. 2b). Furthermore, the feature importance plot indicated that "Diameter mean" was the most significant predictor for density estimation (see Fig. 2c). The feature interaction heatmap showed that the combination of "Diameter mean" and "Height mean" exhibited the strongest interaction intensity, making it the most crucial pairing in our analysis (see Fig. 2d). The SHapley Additive exPlanations (SHAP) summary plot clarified that higher values of "Diameter mean" consistently resulted in positive effects on density predictions.[30]

In contrast, "Height std" displayed bidirectional effects (see Fig. 2e). When constructing the PIML model to convert QD size to density, we selected the core input parameters: "Diameter mean", "Diameter std", "Height mean", and "Height std". We screened and compared several regression models based on their performance, ultimately choosing the CatBoost model (see Supplementary Information for the comparison results of different regression models and the performances of the CatBoost model, S5 and S6). Using the QD size data obtained from the first stage of the PIML model, the inference from the CatBoost model suggests that the required QD density ranges from $4.4\times 10^{10}$ $cm^{-2}$ to $5.5 \times 10^{10}$ $cm^{-2}$ as the standard deviation increases.[31]

**RHEED data acquisition and feature analysis**

During growth, the electron beam from the RHEED gun interacts with the continuously rotating sample, producing a diffracted pattern on the fluorescent screen (see Fig. 2f). A reminder is triggered to ensure adequate time to manually open the RHEED fluorescent screen shutter before capturing and preprocessing real-time RHEED images, which are then sent to models for analysis. The preprocessing methods used in this study align with previously reported methods.[13] Each frame of the RHEED data is processed as a single channel of luminance information. These processed data are then stacked along an additional dimension to form a three-dimensional array, which serves as the input sample for the models. Thus, RHEED data are continuously recorded and analyzed in real-time, enabling precise determination of the critical transition temperature. While manual adjustments to the RHEED power supply and the fluorescent screen shutter were necessary in this study, these processes could be improved for computer control by integrating an RHEED system with the optional "microprocessor-controlled beam current regulation" feature and upgrading the fluorescent screen shutter to an automatic control system.

Gathering as many temperature points as possible is essential to obtain the parameter initialization curve. SemiEpi focuses on several vital temperatures, such as the deoxidation temperature, the transition temperature of As cap/c(4×4), c(4×4)/(2×4), and (2×4)/(n×6). We reported RHEED characteristics of GaAs deoxidation, corresponding to the theoretical temperature of 580 °C.[32,33] After deoxidation, the substrate temperature is reduced and subsequently increased to observe the transition from the As cap to the (n×6) stage.[34,35] By observing the brightness and spacing of RHEED streaks from different angles, the ×4 reconstruction line can be identified from two angles, corresponding to c(4×4) (see Fig. 2g-h). As the temperature increases, thermal vibrations of surface atoms intensify, leading to atomic reconstructions. Consequently, a gradual transition from c(4×4) to (2×4) can be observed around the theoretical temperature of 510 °C (see Fig. 2i-j).[26] When the temperature exceeds a theoretical temperature of 620 °C, the (2×4) becomes increasingly blurred, and the (n×6) gradually appears due to the lattice expansion of the GaAs crystal.[24,36] In this study, ×2 and ×6 reconstruction line can be observed from two angles, corresponding to the (2×6) structure (see Fig. 2k-l).

We used a dataset of 38 samples to develop the Initialization Model and Growth Models. Of these, 8 samples were GaAs substrates observed at different temperatures. Since the GaAs surface undergoes a gradual transformation, we collected data from each sample multiple times, with each video lasting at least 20 minutes. The remaining 30 samples were InAs QDs grown under various conditions. The growth for each of these samples was repeated multiple times, with each video recording lasting at least 90 seconds. The dataset used for model training consists of various RHEED images, including clear and blurred examples. During data preprocessing, we applied image augmentation techniques, including adjustments to brightness, saturation, and contrast, along with the noise addition to simulate various levels of blurriness and effects that could result from manual adjustments to the RHEED power supply.[37] Additionally, we utilized random image cropping to mimic the effects of varying beam positions on the sample, which influences the range of patterns displayed. We then converted images into single-channel intensity images for model input.[38] This approach enabled the model to effectively process more degraded images during validation, enhancing its generalizability and resulting in approximately 370,000 NumPy arrays.

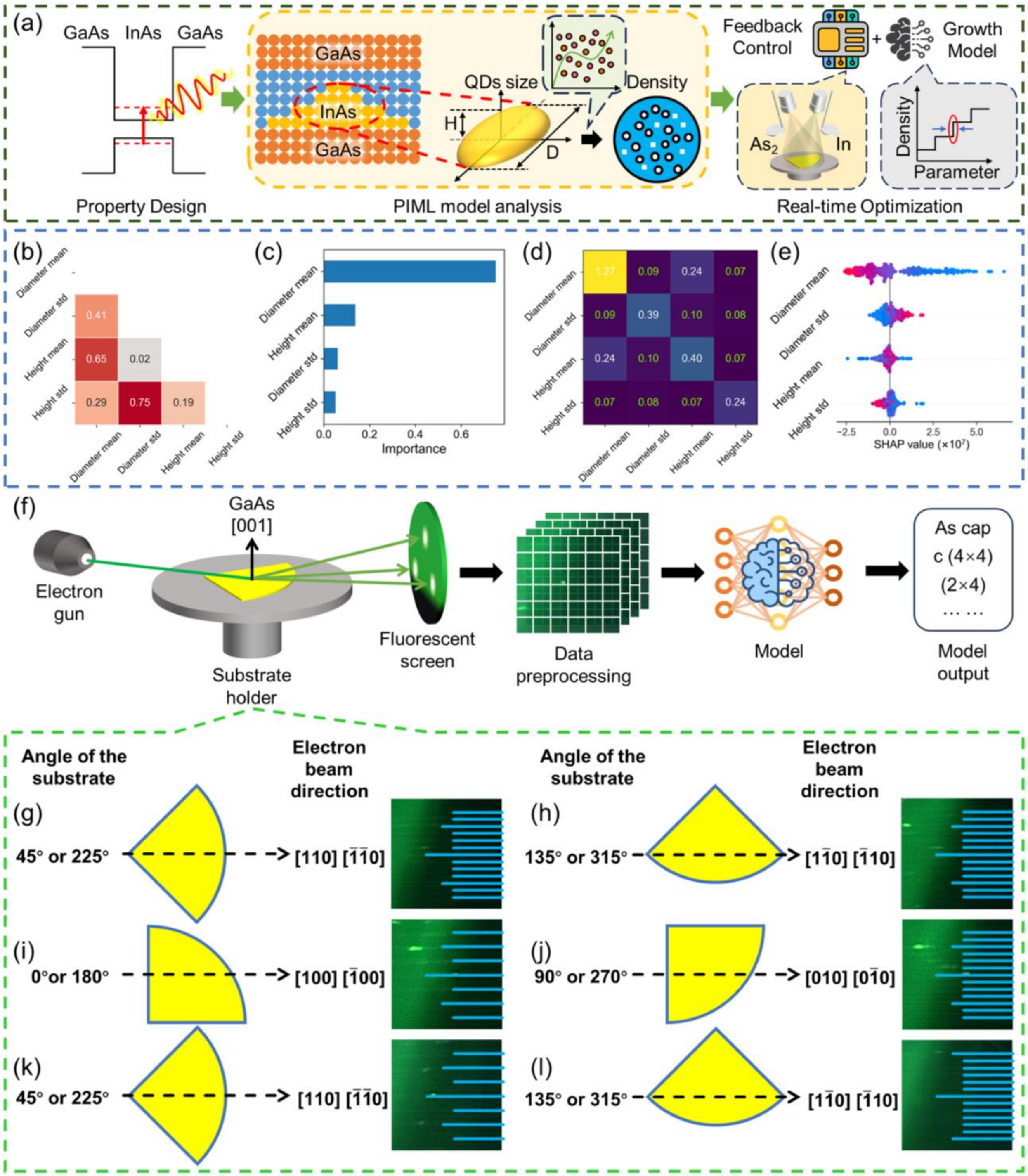

**Fig. 2: The QDs size feature analysis results and RHEED data process methods.** (a) The process to design and grow QDs with emission wavelengths. (b-e) The feature correlation, feature importance, feature interaction heat map, and SHAP summary plot of QD size data. (f) The workflow for RHEED data acquisition and processing. (g-i) Typical RHEED images obtained at various substrate angles. The blue lines on the images represent reconstruction streaks. Longer solid blue lines indicate integer order streaks, while shorter ones denote fractional order streaks.

**Initialization and Growth Models' construction and evaluation**

The conventional convolutional neural network (CNN) approach processes all color channels in an image simultaneously, integrating information from each channel to extract features. This is effective for images with rich color information.[39] However, it is less practical for datasets comprising multiple stacks of single-channel luminance information, such as RHEED data acquired from different angles of a continuously rotating substrate. To address this issue, a model must emphasize channel information, strengthen inter-channel correlations, and enhance data processing efficiency.[40] Therefore, we introduce the global attention residual network (GARN) block and the cross-layer adaptive fusion (CAF) block, which are specifically developed to prioritize channel information. This study constructs three different models to analyze RHEED data of different material growth steps, namely, the "Initialization Model" and two "Growth Models", which are the "Temperature Model" and the "Shutter Model" (see Fig. 3a). These models are designed for a multi-class classification task, with "Initialization Model" identifying reconstruction and deoxidation states, "Temperature Model" predicting the optimal InAs growth temperature, and "Shutter Model" evaluating the completion conditions for InAs QD growth.

The Initialization Model and Growth Models are constructed using GARN block, CAF block, and multilayer perceptron (MLP). GARN block first maps input features to a reduced number of channels through convolutional layers (see Fig. 3b).[41,42] Several generalized attention residual blocks follow this, each comprising two convolutional layers and a channel attention mechanism which adjusts the contribution of each channel by globally pooling the feature map and calculating importance weights. Finally, a downsampling layer and an additional convolutional layer further process the feature map, reducing its spatial resolution to generate the final output, which is a three-dimensional array. The CAF block leverages a transformer-based architecture to process image data (see Fig. 3c).[43,44] The input of a three-dimensional array from GARN is divided into small blocks, and their features are embedded through linear transformations. The transformer's attention mechanism captures global dependencies between image blocks and processes cross-layer associations. Traditional CNN can lead to the loss of vital angle-dependent diffraction details. In contrast, the CAF mechanism excels at processing RHEED images by dynamically adjusting feature channel weights, where essential diffraction information shifts at various angles due to substrate rotation and wobbling. This mechanism allows the CAF to effectively model contextual relationships between different image blocks, enhancing the recognition of complex patterns.[45] The model's feed-forward network further processes these features, which are finally classified using a MLP. This adjustment enables the GARN and CAF blocks to identify and extract critical features more accurately, a crucial capability for detecting subtle feature changes in images captured at varying angles.

The number of stacks in the Initialization Model and Growth Models is crucial for delivering richer input, especially when analyzing RHEED data, which experiences periodic changes. Not all frames of RHEED images provide informative data across all angles. If the number of image stacks is too small, the training samples may include significant irrelevant or non-informative data. This can lead to inconsistent training outcomes, unstable validation performance, and even overfitting. Therefore, we optimized the model input sizes and determined that the best training performance was achieved using 24 images per batch, each with a resolution of 128 × 128 pixels (see Fig. 3d-e). The accuracy is calculated as the ratio of correctly predicted labels to the total number of predictions. In our work, we selected the cross-entropy loss function, a well-established choice for classification tasks. This function guides the model in adjusting its weights and biases by minimizing the loss, thereby enhancing the precision of predictions.[46,47] To expedite the training process, we employed the stochastic gradient descent optimizer, which effectively updates the model parameters during each iteration.[48,49] The resulting validation accuracies for the “Initialization Model”, “Temperature Model”, and “Shutter Model” reached 99.1%, 99.6%, and 99.9%, respectively.

We analyzed alignment features of the “Initialization Model”. Using t-Distributed Stochastic Neighbor Embedding (t-SNE) analysis, we observed a clear separation between different color points, indicating the model effectively identifies both deoxidation and various reconstruction states (see Fig. 3f).[50] We also extracted two frames of typical RHEED maps from a sample, both showing prominent ×2 (see Fig. 3g-h). By plotting the convolutional features map, we observed that the convolution primarily focuses on streak-like features in RHEED images, which correspond to different reconstruction streaks, demonstrating high interpretability (see Fig. 3i-j). When analyzing the attention heat map of the model with these two typical RHEED frames as inputs, it is evident that non-streak features appear dark, indicating that the model did not focus on these regions (see Fig. 3k-l). Conversely, the brighter regions highlight streak-shaped features. Feature maps were also combined from multiple images to identify focal regions across the model. These plots highlight streak features on the left side (see Fig. 3m). In contrast, the right side, consisting of non-streak features, is mainly flat. This indicates the model effectively focuses on regions with distinct features. Additionally, we used Gradient-weighted Class Activation Mapping (Grad-CAM) to analyze the contribution of each region to the classification results (see Fig. 3n).[51-53] The Grad-CAM results highlight multiple streak features, which align well with the streak feature locations in the input image, demonstrating that the model has high sensitivity in data processing.

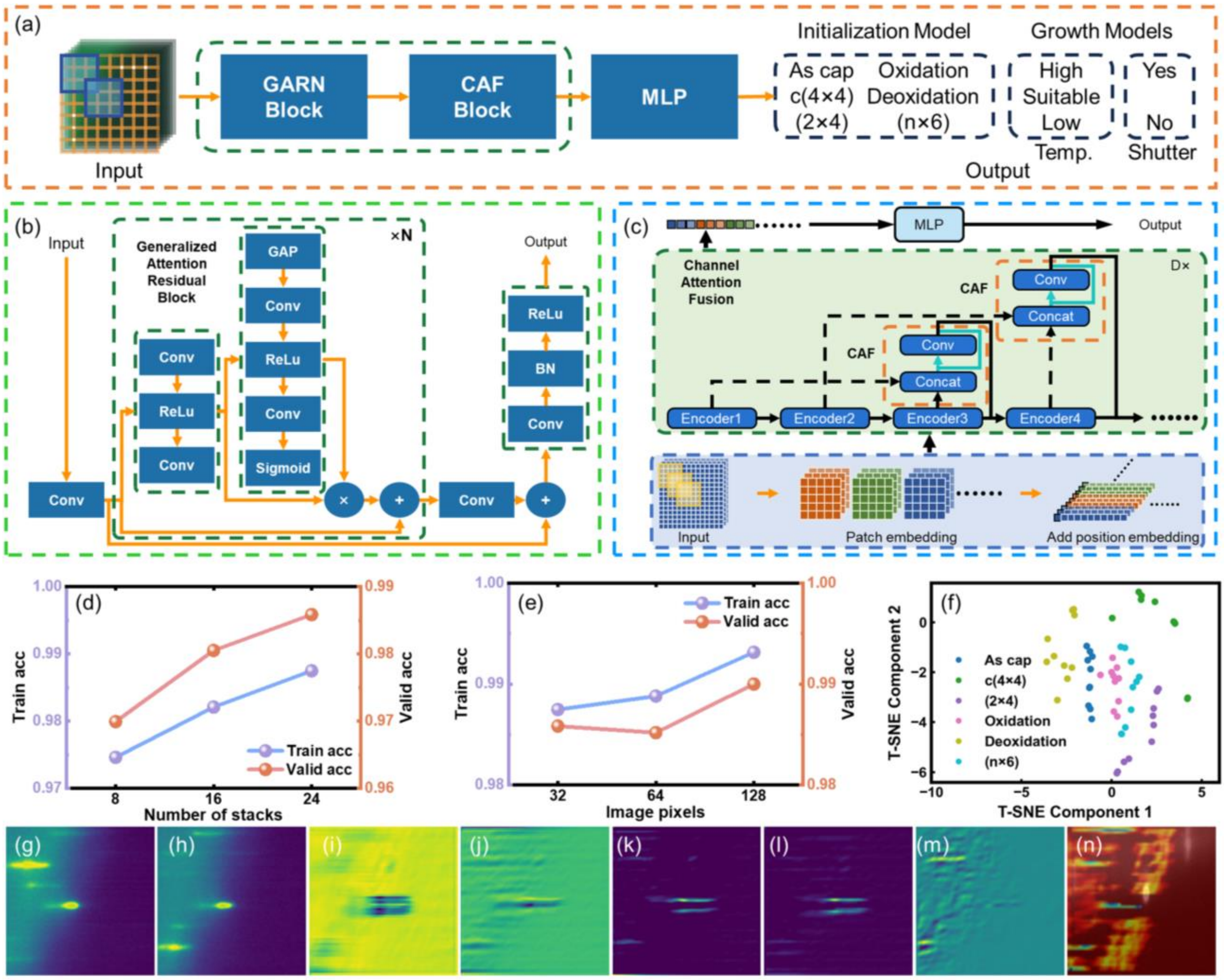

**Fig. 3: Initialization and Growth Models' construction and evaluation.** (a) A simplified architectural diagram of the construction of the "Initialization Model" and "Growth Models". The structure of the (b) GARN and (c) CFA blocks. Conv: convolutional layer. ReLu: rectified linear unit activation layer. GAP: global average pooling layer. Sigmoid: sigmoid activation layer. BN: batch normalization layer. Encoder: transformer encoder. MLP: multi-layer perceptrons. The variation of model validation accuracy and validation loss under different (d) heads and depths and (e) image stacks. Model feature processing: (f) t-SNE visualization of high-dimensional

features. (g-h) Typical RHEED images. (i-j) Convolutional layer feature maps, (k-l) Attention heat maps, (m) Combined feature map, (n) Gradient-weighted class activation mapping results.

**Data labeling**

The dataset used to train the "Initialization Model" was collected through real-time RHEED imaging of a GaAs substrate during its deoxidation process, as well as from various reconstruction states observed at different temperatures during the heating of the smooth GaAs surface, including the "Oxidation", "Deoxidation", As cap, c(4×4), (2×4), and (n×6) phases. These stages were recorded to create the dataset for training the model and labeled accordingly. Data corresponding to intermediate surface reconstruction stages were excluded from the dataset due to the challenges in assigning clear labels to these transitional states. As a result, the model recognizes a successful transformation of the material's surface when it consistently identifies a newly reconstructed state multiple times in succession.

We utilized data from InAs QDs growth under various conditions to train the "Temperature Model" and "Shutter Model". By characterizing the QD density using AFM, we analyzed how different growth temperatures affected its density, enabling us to design temperature regulation strategies for achieving the desired density of QDs. The analysis revealed that higher substrate temperatures during growth facilitate the formation of low-density QDs, whereas lower temperatures favor high-density QDs. To enhance the generalization capability of our classification model, particularly for samples near the decision boundary where label overlap may occur, we adopt a broader QD density range. This expanded range encompasses the density values predicted by the PIML model and targeted during material growth. Moreover, using a wider range of density labels better reflects the natural fluctuations observed in practice and helps reduce validation errors in QD density measurements obtained from AFM, thereby improving the overall robustness of the model. We observed that the density of InAs QDs for a deposition amount of 2.6 ML ranges between $4 \times 10^{10}$ $cm^{-2}$ and $6 \times 10^{10}$ $cm^{-2}$.[54,55] Therefore, within the "Temperature Model", RHEED data from samples with densities within this density range were categorized as "Suitable". Data from samples with densities below $4 \times 10^{10}$ $cm^{-2}$ were categorized as "High". If the model output shows "High", the substrate temperature is too high and should be decreased. Conversely, data from samples with densities above $6 \times 10^{10}$ $cm^{-2}$ were labeled "Low". If the model output indicates "Low", this indicates the substrate temperature is too low and should be increased. This mapping of model-generated labels to temperatures enables SemiEpi to provide optimized guidance throughout the growth process.

To achieve the desired QD density accurately, precise control of the growth temperature and timely completion of the growth process is essential. RHEED images captured facilitate the identification of the growth stages both before and after the InAs QD formation, which provided the necessary labels for the "Shutter Model". The RHEED patterns before and 10 seconds after the end of growth of samples with QD densities in the range of $4\text{-}6 \times 10^{10}$ $cm^{-2}$ were categorized as "Yes", while those outside this range were categorized as "No". This approach ensures the timely closure of the In shutter when the QD density reaches $4\text{-}6 \times 10^{10}$ $cm^{-2}$.

The "Initialization Model" and the "Shutter Model" are discriminative models that evaluate the material's state, identifies and optimizes the InAs QD growth conditions in real time. In contrast, the "Temperature Model" is a predictive model that forecasts the performance of samples upon growth completion. This model facilitates real-time predictions of QD density during growth, enabling timely parameter adjustments and ensuring high-quality material growth.

## Experiment validation

### Parameter initialization

In our study, thermocouples are used to measure temperatures; however, temperature readings from multiple reconstruction transition points exhibit a non-linear relationship with the corresponding theoretical temperatures. A curve fitting approach is necessary to achieve a more reliable correlation.[56,57] We compared different curve-fitting methods and found that the quadratic curve-fitting results were better than those obtained with linear or higher-order polynomial fits (see Supplementary Information for the comparison results from different function fits, S7).

The parameter initialization curve extracts the thermocouple temperatures corresponding to the growth conditions to achieve the desired QD density, which can be achieved at a growth theoretical temperature of 490 °C with a low growth rate. While it is possible to optimize the growth temperature for GaAs during the process, variations around the theoretical temperature of 600 °C have minimal impact on the quality of the material (see Supplementary Information for the surface morphology and crystal quality of GaAs grown at different temperatures, S8). Therefore, with the parameter initialization curve, the thermocouple temperatures corresponding to the theoretical temperature of 490 °C and 600 °C were established for the growth of InAs QDs and GaAs.

The main objective of SemiEpi is to generate parameter initialization curves and optimize growth parameters for InAs QDs with desired wavelength. Real-time data recorded by SemiEpi around different transition points was analyzed (see Fig. 4). During the automatic deoxidation process, SemiEpi heated up in increments of 5 °C, from the thermocouple temperature of 390 °C to 415 °C through five increments (see Fig. 4a). The RHEED screen showed no distinctive features until the thermocouple temperature of 415 °C, when distinct bright spot features appeared (see Fig. 4c-d). Analysis of the "Initialization Model" output during deoxidation revealed that, from the $0^{th}$ to around the 16,000$^{th}$ sequence, the oxidation probability remained close to 1, indicating that the model did not identify the deoxidation state. After the 16,000$^{th}$ sequence, the deoxidation probability rapidly increased to nearly 1, confirming the model's accurate identification of the deoxidation state (see Fig. 4b).

After growing a layer of GaAs on the deoxidized sample and cooling, the substrate was gradually heated at a rate of 15 °C per minute (see Fig. 4e). Real-time RHEED data collected during this process was analyzed using the "Initialization Model" to monitor reconstruction states (see Fig. 4f). Initially, the model identified only "As cap" labels, with RHEED patterns showing diffuse features from the beginning to around the 2,000$^{th}$ sequence (see Fig. 4g-h). By around the 3,000$^{th}$ sequence, at a thermocouple temperature of 135 °C, a marked shift occurred as the probability of "As cap" labels dropped sharply and "c(4×4)" labels became dominant. The RHEED patterns began exhibiting periodic features consistent with the emergence of a ×4 reconstruction (see Fig. 4i-j). At around the 11,000$^{th}$ sequence, as the substrate heating reached 353°C, the probability of "(2×4)" labels increased significantly. This shift indicated a structural change on the surface, supported by RHEED observations of features combining ×2 and ×4 (see Fig. 4k-l). Finally, by the 14,000$^{th}$ sequence, the model output showed an increased probability for the "(n×6)" label. At this stage, the thermocouple temperature reached 455 °C, and the RHEED patterns displayed a distinct ×6 periodicity alongside residual ×2 features (see Fig. 4m-n). The results demonstrate the model's high sensitivity in distinguishing between different structural states.

Finally, SemiEpi used a quadratic curve to fit the collected thermocouple temperature data to theoretical temperature data. This sets the initial thermocouple temperatures for InAs growth at 330 °C and for GaAs growth at 436 °C (see Fig. 4o).

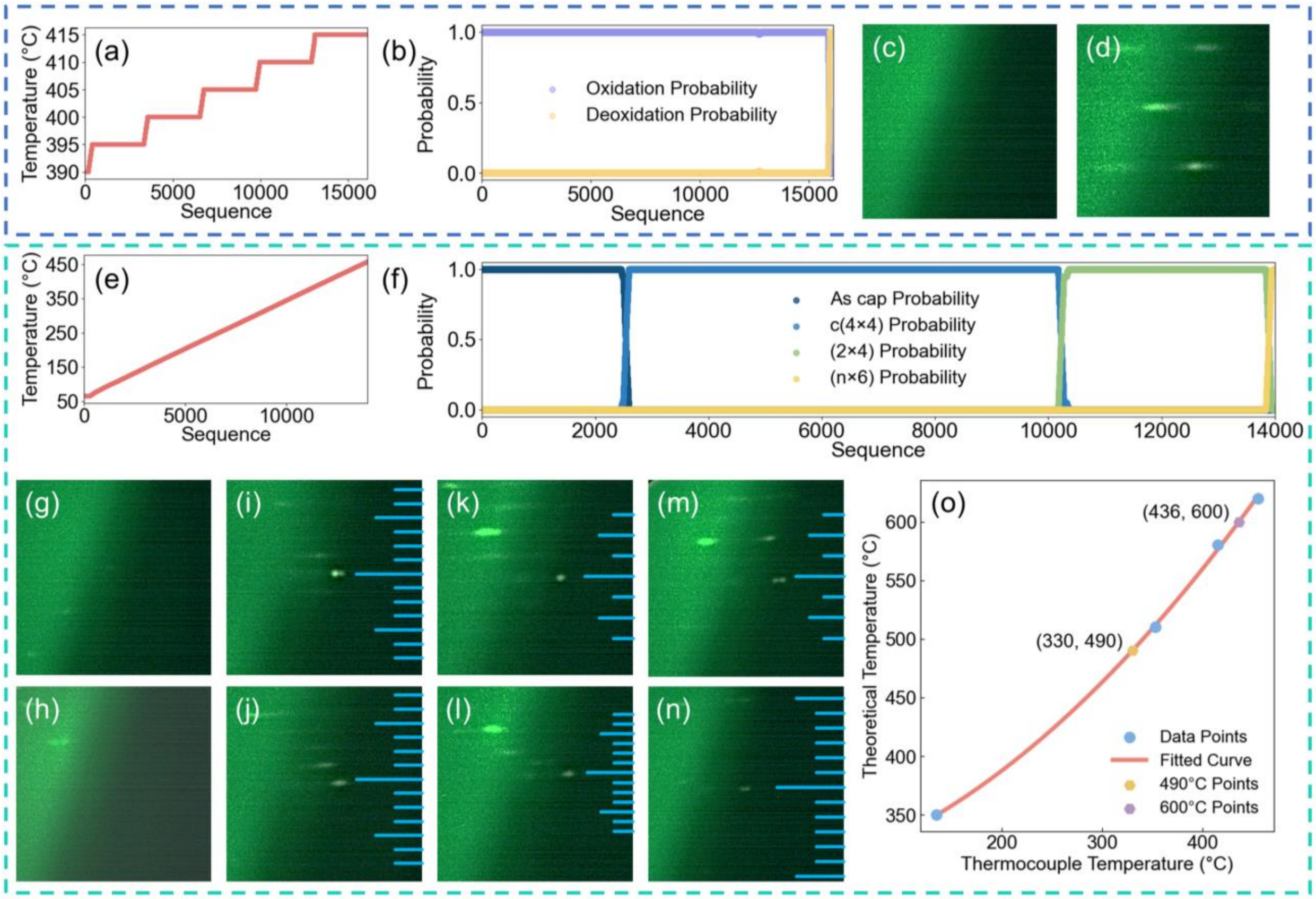

**Fig. 4: The parameter initialization.** (a) Substrate temperature and (b) the running average results of "Initialization Model" output during substrate deoxidation. The RHEED image captured at around (c) 13,000$^{th}$ and (d) 16,000$^{th}$ sequence of (a). (e) Substrate temperature and (f) the running average results of "Initialization Model" output during parameter initialization. The RHEED image captured at around (g-h) 2,000$^{th}$, (i-j) 3,000$^{th}$, (k-l) 11,000$^{th}$, and (m-n) 14,000$^{th}$ sequence of (e). RHEED images were captured from two angles. (o) Parameter initialization results. Source data are provided as a Source Data file.

## QD growth

The density of InAs QDs is highly sensitive to temperatures. Based on the output of the PIML model, SemiEpi uses RHEED to analyze the material surface and optimize the temperature in real-time to ensure that the samples achieve the desired density range of 4-6 × $10^{10}$ $cm^{-2}$ (see Fig. 5) (see Supplementary Video for the experiment).[58,59] During the growth of buried InAs QDs, the thermocouple temperature analysis revealed a 2 °C decrease, consistent with the trend of QD density change relative to thermocouple temperature (see Fig. 5a and Supplementary Information for the variation in QD density with thermocouple temperature, S9). This indicates that the initial temperature alone does not ensure the desired density. Statistical analysis of RHEED data using the "Temperature Model" reveals that, in the early stages, the model predominantly outputs the "High" label (see Fig. 5b). As growth progresses, the probability of the model outputting the "Suitable" label increases around the 200$^{th}$ sequence. Subsequently, the model mainly outputs "Suitable" labels until the end of the growth. The probability of the "Low" label remains consistently low, indicating the initial growth temperature was too high and required adjustment to achieve the desired density.

The "Shutter Model" analysis indicates that from the 0$^{th}$ to approximately the 1200$^{th}$ sequence, the model primarily outputs "No", suggesting the InAs QD growth has not yet achieved the desired density (see Fig. 5c). However, at the 1300$^{th}$ sequence, the probability of outputting "Yes" gradually increases, indicating the desire QD density is achieved. Analysis of RHEED data reveals that from the 100$^{th}$ and 700$^{th}$ sequences, RHEED patterns exhibit evident streak characteristics with no QDs formed (see Fig. 5d-e). In contrast, RHEED patterns from the 1300$^{th}$ sequence reveal well-formed, rounded, and neatly aligned spots, indicating successful QD formation (see Fig. 5f).

Additionally, the sample prepared using SemiEpi exhibited a photoluminescence (PL) intensity of 3605 and a full width at half maximum (FWHM) of 29.13 meV at the wavelength of 1235 nm. This performance significantly surpassed that of the reference sample prepared using conventional methods, which exhibited a PL intensity of 2270 and an FWHM of 34.70 meV at the wavelength of 1229 nm. Additionally, it outperformed the reference sample prepared using parameter initialization, which exhibited a PL intensity of 2531 and an FWHM of 30.79 meV at the wavelength of 1225 nm (see Fig. 5g and Supplementary Information for the reference sample prepared using conventional methods and the reference sample prepared using parameter initialization, S10 and S11). The obtained effective lifetimes of 2.06 ns from time-resolved photoluminescence (TRPL) measurements demonstrate the high quality of the QDs prepared by SemiEpi, providing strong evidence for the reliability of the platform (see Supplementary Information for the time-resolved photoluminescence results, S12). It is worth noting that the emission wavelengths of samples prepared with the SemiEpi closely match the desired wavelength of 1240 nm. In contrast, the two reference samples exhibit greater deviations from the desired emission wavelengths.

SemiEpi also grew surface InAs QDs to observe their morphology. The initial growth temperature for these QDs was set to the thermocouple temperature of 328 °C, which was the temperature for buried InAs QDs (see Fig. 5h). During growth, the thermocouple temperature remained stable at 328 °C. This highlights that adjusting the temperature during the growth of buried InAs QDs allowed achieving the desired QD density without the need for additional temperature adjustments. The probabilistic statistical analysis of the "Temperature Model" output during the process shows that initially, the model predominantly outputs "Suitable" labels, with only a few "High" labels (see Fig. 5i). This indicates that the initial growth temperature is appropriate. The analysis of the "Shutter Model" output reveals a significant increase in the probability of the "Yes" label at the $1300^{th}$ sequence, indicating that the QDs have achieved the desired QD density (see Fig. 5j).

In the RHEED patterns obtained at the $100^{th}$, $700^{th}$, and $1300^{th}$ sequences, we observe a transition from streak to spot features as the growth progresses (see Fig. 5k-m). The spot pattern at the $1300^{th}$ sequence appears well-ordered and rounded. AFM characterization of the surface InAs QDs indicates a high QD density of approximately $5\times10^{10}$ $cm^{-2}$, with a relatively uniform distribution (see Fig. 5n). In addition, the reference sample prepared using parameter initialization achieved a density of $5.4 \times 10^{10}$ $cm^{-2}$. However, the reference sample prepared using conventional methods only achieved a QD density of $3.7 \times 10^{10}$ $cm^{-2}$, falling below the range designated as "Suitable". This highlights the effectiveness of parameter initialization in improving growth outcomes.

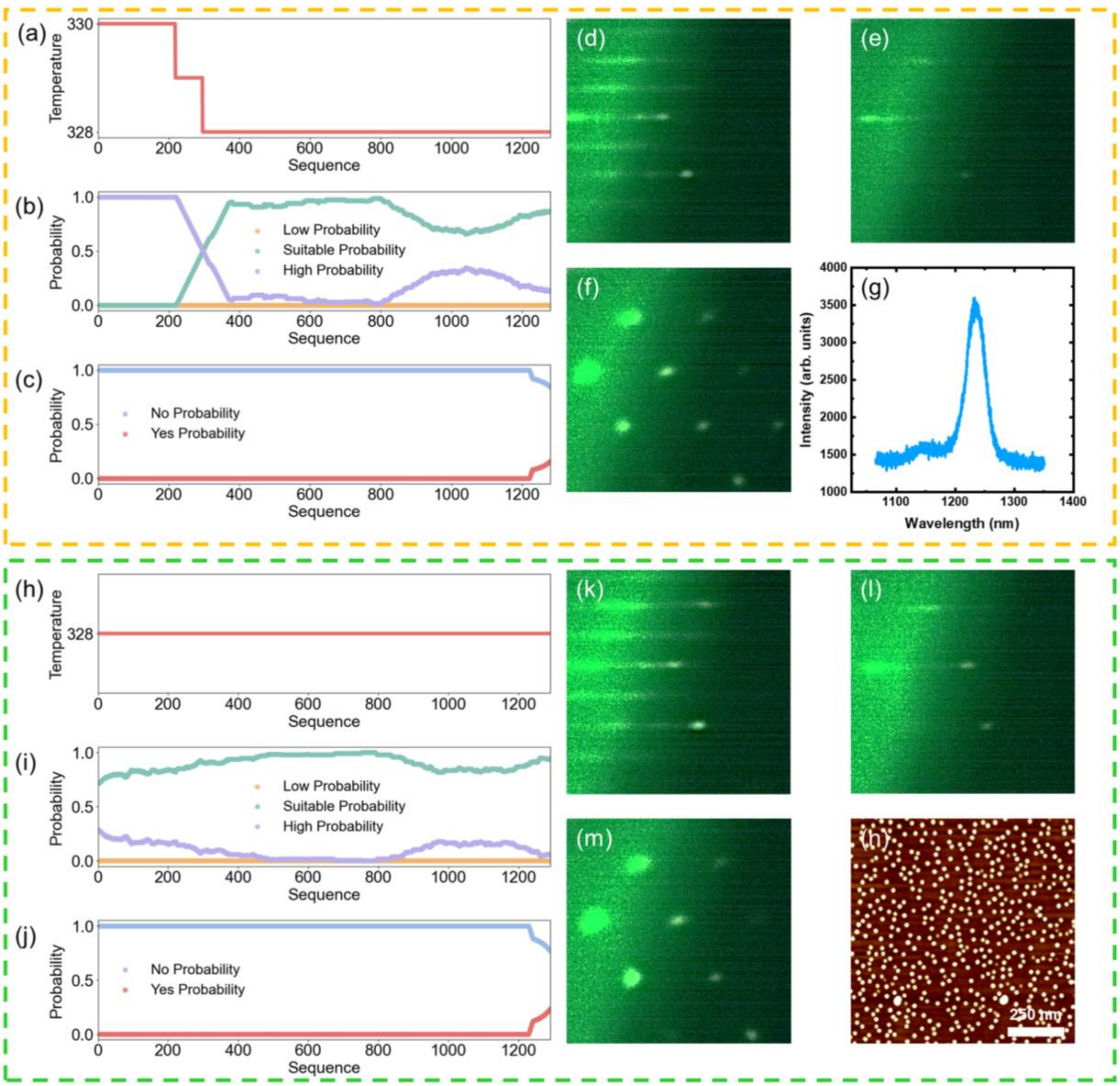

**Fig. 5: The growth of the InAs QDs.** (a) Substrate temperature, and the running average results of (b) "Temperature Model" and (c) "Shutter Model" output during growth buried InAs QDs. The RHEED image captured at around (d)$100^{th}$, (e) $700^{th}$, and (f) $1300^{th}$ sequence of (a). (g) PL spectrum of the buried QDs. (h) Substrate temperature, and the running average results of (i) "Temperature Model" and (j) "Shutter Model" output during growth surface InAs QDs. The RHEED image captured at around (k)$100^{th}$, (l) $700^{th}$, and (m) $1300^{th}$ sequence of (h). (n) The 1 μm × 1 μm AFM image of the surface QDs. Source data are provided as a Source Data file.

## Discussion

In this work, we present a self-driving approach for on-demand heterostructure growth. We introduce SemiEpi, a system that designs heterostructures using PIML models while leveraging the substrate's intrinsic properties to establish optimal initial conditions. These conditions are proposed for experimental validation by aligning the thermocouple temperature with theoretical values through parameter initialization. It continuously monitors and analyzes these conditions using ML algorithms to predict outcomes, dynamically adjusting them throughout the InAs QDs growth process. This method effectively maps the relationships between material status and their outcomes, enhancing material quality and ensuring consistent and reliable results.

The QDs' performance using SemiEpi aligns with the expected characteristics of QD lasers.[60-62] However, our experiments are facing some challenges, including a failure rate associated with temperature change rates and the similarity of data characteristics. The similarity in RHEED patterns between the As cap layer and the oxidation state complicates the identification process and impacts the dataset construction. This emphasizes the need for more accurate labeling and clearer differentiation between these states.

In summary, SemiEpi integrates PIML models, prior semiconductor knowledge, and real-time feedback control during experiments, offering a robust framework to address hardware disparities between systems-a key source of unpredictable results in semiconductor research. The field faces inherent data limitations due to the high cost and prolonged turnaround. SemiEpi overcomes the limitations of traditional methods, which rely on extensive datasets and fixed programs while being highly sensitive to reactor geometry, operational status, and sample inconsistencies. By identifying growth stages and minimizing cross-reactor variability through parameter initialization, the platform ensures reproducible growth conditions. We utilized RHEED for analysis, but the system is readily adaptable to other in-situ techniques, such as absorption spectroscopy and mass spectrometry. By dynamic parameter adjustments, SemiEpi effectively explores and optimizes the parameter space linking in-situ data and material growth results, ensuring that growth conditions remain stable and reproducible, regardless of external fluctuations.

We utilize existing physical laws to design sizes that achieve the desired emission wavelengths. Additionally, we apply ML to uncover high-dimensional, nonlinear relationships to establish correlations between size and density when the physical laws for these variables are not available. This approach can be generalized to other scenarios. Moreover, our method is scalable for large-scale material production, significantly reducing optimization cycles and improving yield quality. Moving forward, integrating portability, multimodality, and reinforcement learning into smart MBE systems could transform the future of precision material growth.

## Materials and Methods

### Material growth

The InAs QD samples were grown on GaAs substrates using a Riber 32P MBE reactor. The system is equipped with an arsenic (As) valved cracker, as well as indium (In) and gallium (Ga) effusion cells. During the growth process, the $As_2$ source was used, and the cracker temperature was maintained at over 900 °C. Beam Equivalent Pressure (BEP) measurements were used to assess the fluxes and calibrate the ratios of group III and V elements.[63] Substrate temperatures were monitored using Type-C thermocouples, and the growth rates were calibrated by observing RHEED oscillations from additional layers grown on the GaAs substrate. The BEP values for the cells were as follows: $6.8 \times 10^{-9}$ Torr for In, $1.5 \times 10^{-7}$ Torr and $9 \times 10^{-8}$ Torr for Ga, and $2.5 \times 10^{-6}$ Torr, $1.5 \times 10^{-6}$ Torr, and $1.0 \times 10^{-6}$ Torr for As used in GaAs, and InAs, respectively. Before the growth process, the n-GaAs substrates were outgassed in a buffer chamber at 350°C. The growth was managed by SemiEpi, which included steps for substrate deoxidation, parameter initialization, and material growth. The growth rates were 0.6 μm/h and 0.36 μm/h for GaAs, while the InAs growth rate was approximately 0.016 ML/s.

**Material characterization**

RHEED was set up in the MBE growth chamber with a power supply at 12 kV and 1.49 A to generate an electron beam (RHEED 12 from STAIB Instruments). The electron beam interacts with the surface of the epitaxial layer, producing diffraction patterns that are then projected onto a fluorescent screen. These patterns are captured in real-time via a camera mounted outside the chamber in a dark room. Throughout the growth process, the substrate was rotated at one revolution every 3 seconds. The camera had an exposure time of 100 ms and a sampling rate of 8 frames per second, meaning that every 24 frames represented one full revolution of the substrate. After the growth was completed, the samples were characterized using a custom-built PL system (iHR 550 spectrometers from HORIBA). This system consists of an optical beam splitter, reflector, attenuator, and a 532 nm continuous wave excitation laser. An InGaAs detector within the spectrometer collected the light emitted by the samples. The surface morphology of the InAs QDs was also characterized using AFM (Dimension Icon from Burker). Time-resolved PL (TRPL) measurements were conducted at 130 K using a short-pulse laser (Chameleon Ultra from Coherent). The excitation power density was set at 0.76 W/cm. The emitted PL was analyzed with a spectrometer (FHR 1000 from Horiba) and detected using an infrared single-photon detector (ID230 Infrared Single-Photon Detector from Picoquant).

## Data Availability

Source data are provided with this paper.

## Code Availability

Source codes are provided within this paper. The ML model and LabVIEW program used can be found via GitHub at https://github.com/SEMIML/GARN-CAF-model-QDs.

## Acknowledgements

This work was supported by Beijing Natural Science Foundation (Grant No. F251036 and L248103, C. Z.), the National Key R&D Program of China (Grant No. 2023YFB2805900, S. M. C.), National Natural Science Foundation of China (Grant No. 62274159, C. Z.; 62204072, J. N. D), and CAS Project for Young Scientists in Basic Research (Grant No. YSBR-056, C. Z.).

**Author Contributions Statement**

C. Z. conceived the idea, designed the investigations and the growth experiments. C. S., W. K. Z., and B. X. performed the molecular beam epitaxial growth. C. S., Y. L., and F. X. L. performed the model construction and computation. C. S., S. J. P, K. Y. X., H. C., C. X., X. T. C, R. X. L, Z. B. N., and C. Y. J did the sample characterization. C. S., S. M. C, Z. M. W, C. L. X., and C. Z. wrote the manuscript. C. Z. led the molecular beam epitaxy program. Z. G. W. supervised the team. All authors have read, contributed to, and approved the final version of the manuscript.

**Competing Interests Statement**

The authors declare no competing interests.

**Supplementary Information**

Details on the platform environment and program interface, the functional interface switch in the program, the deployment environment and hardware wiring scheme, the construction of the physics model, the comparison results of different regression models, the performances of the CatBoost model, the comparison results from different function fits, the surface morphology and crystal quality of GaAs grown at different temperatures, the variation in QD density with thermocouple temperature, the reference sample prepared using conventional methods, the reference sample prepared using parameter initialization, the time-resolved photoluminescence results. The video demonstrated InAs QD heterostructure growth through SemiEpi.

**TOC**

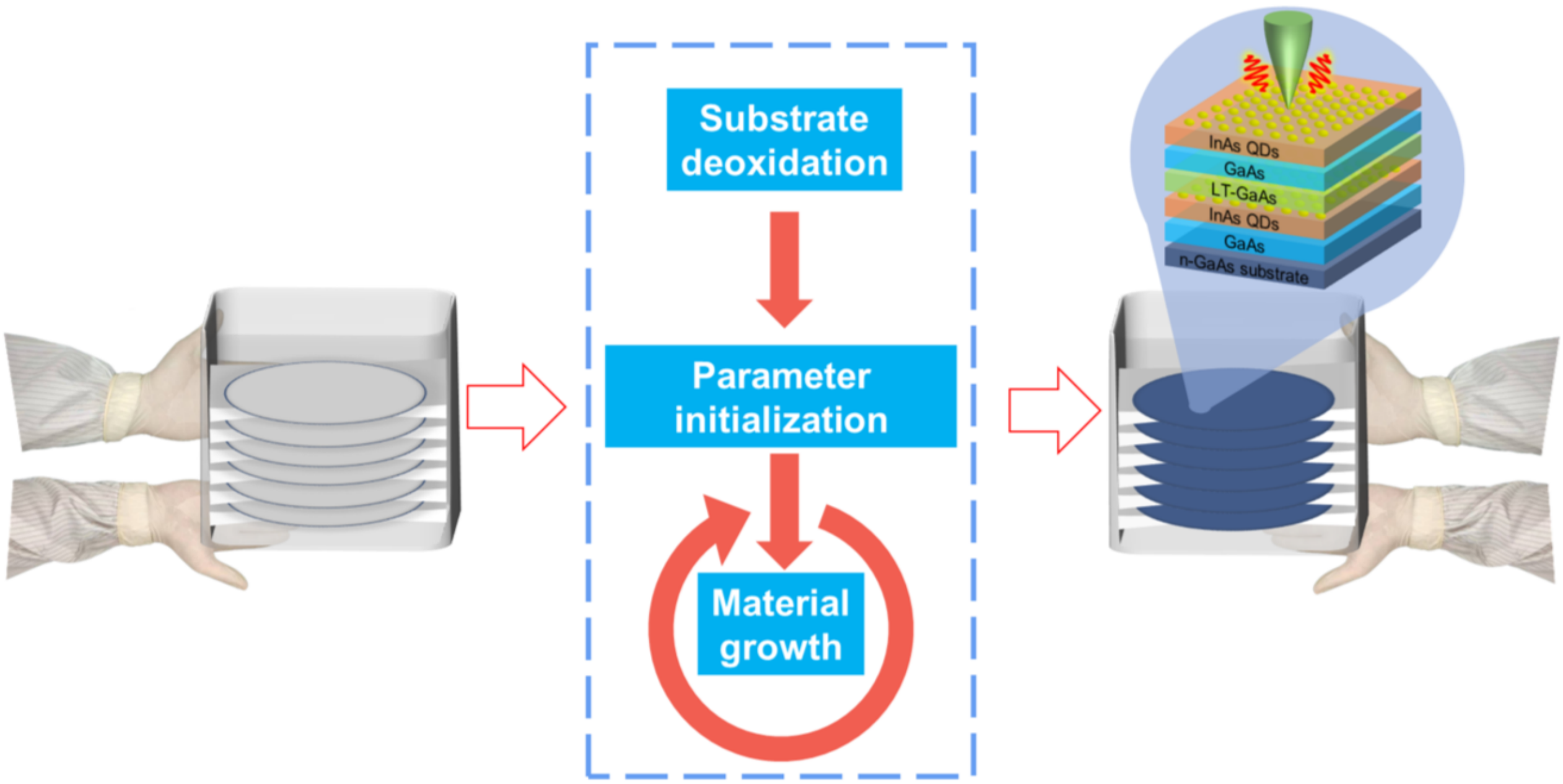